\documentclass[letterpaper]{article} 
\usepackage{aaai25}  
\usepackage{times}  
\usepackage{helvet}  
\usepackage{courier}  
\usepackage[hyphens]{url}  
\usepackage{graphicx} 
\usepackage{natbib}  
\usepackage{caption} 
\usepackage{amsmath}
\usepackage{amsfonts}
\usepackage{booktabs}
\usepackage{siunitx}
\usepackage[table]{xcolor}
\usepackage{makecell}
\usepackage{float}
\usepackage{subcaption}

\usepackage{graphicx,overpic}

\DeclareMathOperator*{\argmin}{arg\,min}

\usepackage{colortbl} 
\definecolor{soft_teal}{HTML}{2FBEAD} 
\definecolor{soft_tan}{HTML}{E8D5C4}

\newcommand{\bc}[1]{\cellcolor{soft_tan}{#1}}

\usepackage{newfloat}
\usepackage{listings}
\DeclareCaptionStyle{ruled}{labelfont=normalfont,labelsep=colon,strut=off} 
\floatstyle{ruled}
\newfloat{listing}{tb}{lst}{}
\floatname{listing}{Listing}
\title{MAPF-GPT: Imitation Learning for Multi-Agent Pathfinding at Scale}
\author {
    Anton Andreychuk\textsuperscript{\rm 1}\equalcontrib,
    Konstantin Yakovlev\textsuperscript{\rm 2,1},
    Aleksandr Panov\textsuperscript{\rm 1,2,3},
    Alexey Skrynnik\textsuperscript{\rm 1,2,3}\equalcontrib
}

\affiliations {
    \textsuperscript{\rm 1}AIRI, Moscow, Russia\\
    \textsuperscript{\rm 2}Federal Research Center ``Computer Science and Control'' of the Russian Academy of Sciences, Moscow, Russia\\
    \textsuperscript{\rm 3}Moscow Institute of Physics and Technology, Dolgoprudny, Russia\\
    andreychuk@airi.net,
    yakovlev@isa.ru,
    panov@airi.net,
    skrynnikalexey@gmail.com
}

\begin{document}

\maketitle

\begin{abstract}

Multi-agent pathfinding (MAPF) is a problem that generally requires finding collision-free paths for multiple agents in a shared environment. Solving MAPF optimally, even under restrictive assumptions, is NP-hard, yet efficient solutions for this problem are critical for numerous applications, such as automated warehouses and transportation systems. Recently, learning-based approaches to MAPF have gained attention, particularly those leveraging deep reinforcement learning. Typically, such learning-based MAPF solvers are augmented with additional components like single-agent planning or communication. Orthogonally, in this work we rely solely on imitation learning that leverages a large dataset of expert MAPF solutions and transformer-based neural network to create a foundation model for MAPF called MAPF-GPT. The latter is capable of generating actions without additional heuristics or communication. MAPF-GPT demonstrates zero-shot learning abilities when solving the MAPF problems that are not present in the training dataset. We show that MAPF-GPT is able to outperform the state-of-the-art learnable MAPF solvers on a diverse range of problem instances and is computationally efficient during inference.


\end{abstract}

\begin{links}
    \small
    \link{Project Page}{https://sites.google.com/view/mapf-gpt/}
\end{links}

\section*{Introduction}

Multi-agent pathfinding (MAPF)~\cite{stern2019multi} is a combinatorial computational problem that asks to find a set of paths for the agents that operate in a shared environment, such that accurately following these paths does not lead to collisions and, preferably, each agent reaches its specified goal as soon as possible. On the one hand, even under simplified assumptions, such as a graph representation of the workspace, discretized time, uniform duration of actions, optimally solving MAPF is NP-Hard~\cite{surynek2010optimization}. On the other hand, efficient MAPF solutions are highly demanded in numerous real-world applications, such as automated warehouses~\cite{li2021lifelong}, railway scheduling~\cite{svancara2022tackling}, transportation systems~\cite{li2023intersection}, etc. This has resulted in a noticeable surge of interest in MAPF and the emergence of a large body of research devoted to this topic. 

Recently, learning-based MAPF solvers have come on stage~\cite{skrynnik2021hybrid,alkazzi2024comprehensive, skrynnik2023switch}. They mostly rely on deep reinforcement learning and typically involve additional components to enhance their performance, such as single-agent planning, inter-agent communication, etc. Meanwhile, in the realm of machine learning, currently, the most impressive progress is driven by self-supervised learning (at scale) on expert data and employing transformer-based architectures~\cite{vaswani2023attentionneed}. It is this combination that recently led to the creation of the seminal large-language models (LLMs) and (large) vision-language models (VLMs) that achieve an unprecedented level of performance in text, image and video generation~\cite{dubey2024llama,liu2024visual,zhuminigpt}. Moreover, such data-driven approach has become widespread in robotics, where an imitation policy is trained based on a variety of expert trajectories~\cite{chen2021decision,chi2023diffusionpolicy}. 
Thus, in this work we are motivated by the following question: \emph{Is it possible to create a strong learnable MAPF solver (that outperforms state-of-the-art competitors) purely on the basis of supervised learning (at scale) on expert data omitting additional decision-aiding routines?} Our answer is positive.

To create our learning-based MAPF solver, which we name MAPF-GPT, we, first, design a vocabulary of terms, called \emph{tokens} in machine learning, that are used to describe any observation an individual agent may perceive and any action it may perform. Next, we create a diverse dataset of expert data, i.e., successful MAPF solutions, utilizing a state-of-the-art MAPF solver. Consequently, we convert these MAPF solutions into sequences of \emph{observation-action} tuples, encoded with our tokens, and utilize a transformer-based non-autoregressive neural network to learn to predict the correct action provided with the observation. In our extensive empirical evaluation, we show that MAPF-GPT notably overpasses the current best-performing learnable-MAPF solvers (without any usage of additional planning or communication mechanisms), especially when it comes to out-of-distribution evaluation, i.e., evaluating the solvers on the problem instances that are not similar to the ones used for training (a common bottleneck for learning-based solvers). We also report ablation studies and evaluate MAPF-GPT in another type of MAPF, i.e. the Lifelong MAPF (both in zero-shot and fine-tuning regimes). 

To summarize, we make the following contributions:
\begin{itemize}
    \item We present the largest MAPF dataset for decision-making, containing 1 billion observation-action pairs.
    
    \item We develop an original tokenization procedure to describe agent observations and use it to create MAPF-GPT, a novel learning-based, decentralized MAPF solver built on a state-of-the-art transformer-based neural network. Trained with imitation learning, MAPF-GPT serves as a foundation model for MAPF tasks, demonstrating zero-shot learning abilities on unseen maps.


    \item We extensively study MAPF-GPT and compare it with state-of-the-art decentralized learning-based approaches, showing MAPF-GPT’s high performance across a wide range of tasks, along with better runtime efficiency.
\end{itemize}

\section*{Related Works}

\paragraph{Multi-agent pathfinding} Several orthogonal approaches to tackle MAPF can be distinguished. First, dedicated rule-based MAPF solvers exist that are tailored to obtaining MAPF solutions fast, yet no bounds on their costs are guaranteed~\cite{okumura2023lacam,li2022mapf}. Second, reduction-based approaches to obtain optimal MAPF solutions are widespread. They convert MAPF to some other well-established computer science problem, e.g., minimum-flow on graphs, boolean satisfiability (SAT), and employ an off-the-shelf solver to obtain the solution of this problem~\cite{yu2013multi,surynek2016efficient}. Next, a plethora of search-based MAPF solvers exist~\cite{sharon2015conflict,sharon2013increasing,Wagner2011}. They explicitly rely on graph-search techniques to obtain MAPF solutions and often may provide certain desirable guarantees, e.g., optimal or bounded suboptimal solutions. Meanwhile, simplistic search-based planners that lack strong guarantees, like prioritized planning~\cite{ma2019searching}, are also widespread.

Recently, learning-based MAPF solvers gained attention. One of the first such successful solvers was PRIMAL~\cite{sartoretti2019primal} that demonstrated how MAPF problem can be solved in a decentralized manner utilizing machine learning. The recent learnable MAPF solvers such as SCRIMP~\cite{wang2023scrimp}, DCC~\cite{ma2021learning}, Follower~\cite{skrynnik2023learn}, to name a few, typically rely on reinforcement learning \emph{and} on additional modules, like the communication one, to solve the problem at hand. Orthogonally to these approaches, we rely purely on imitation learning from expert data.

\paragraph{Offline reinforcement learning} 

Offline deep reinforcement learning develops a policy based on previously collected data without interacting with the environment while training~\cite{levine2020offline}. This allows getting a robust policy due to the utilization of large amounts of pre-collected data. There are numerous effective offline RL approaches, such as CQL~\cite{kumar2020conservative}, IQL~\cite{kostrikov2022offline}, TD3+BC~\cite{fujimoto2021minimalist}. Modern approaches often involve transformers as the architectural backbone. One popular approach is the Decision Transformer (DT)~\cite{chen2021decision}, which models the behavior of an expert by conditioning on the desired outcomes, thereby integrating reward guidance directly into the decision-making process.
In multi-agent scenarios, there is less diversity in offline RL methods; however, a multi-agent adaptation of the DT exists, known as MADT~\cite{meng2021offline}.

\paragraph{Multi-agent imitation learning (MAIL)} Imitation learning and learning from demonstration are actively used in multi-agent systems~\cite{tang2024multi,liu2024learning}. MAIL refers to the problem of agents learning to perform a task in a multi-agent system through observing and imitating expert demonstrations without any knowledge of a reward function from the environment. It has gained particular popularity in the tasks of controlling urban traffic and traffic lights at intersections~\cite{bhattacharyya2018multi,huang2023multi} due to the presence of a large amount of data collected in real conditions and a high-quality simulator (such as Sumo~\cite{lopez2018microscopic}). Among the methods in the field of MAIL, it is possible to note works using the Bayesian approach~\cite{yang2020bayesian}, generative adversarial methods~\cite{song2018multi,li2024gailpg}, statistical tools for capturing multi-agent dependencies~\cite{wang2021multi}, low-rank subspaces~\cite{shih2022conditional}, latent multi-agent coordination models~\cite{le2017coordinated}, decision transformers~\cite{meng2021offline}, etc. Demonstrations are often used for pre-training in game tasks, such as training models for chess~\cite{silver2016mastering, ruoss2024amortized}, and in MAPF tasks, as exemplified by SCRIMP~\cite{wang2023scrimp}. However, despite the listed works in this area, a single foundation model has not yet been proposed, the imitation learning, which already gives high results in multi-agent tasks and does not require an additional stage of online learning in the environment. This is largely due to the complexity of the behavioral multi-agent policies in various tasks (such as in StarCraft~\cite{samvelyan2019starcraft} and traffic control) and the lack of large datasets of expert trajectories, which are necessary for effective training of foundation models. In this regard, the MAPF task is a convenient testbed for investigating transformer foundation models in a multi-agent setting, which will provide additional insights for creating such models in other applications, and our work also provides a large dataset for training MAPF models.

\section*{Background}

\paragraph{Multi-agent pathfinding} The classical variant of the MAPF problem is defined by a tuple $(n, \mathcal{G=(V, E)}, S=\{s_1, ..., s_n|~s_i \in \mathcal{V}\}, G=\{g_1, ..., g_n|~g_i \in \mathcal{V}\})$, where $n$ is the number of agents acting in the shared workspace which is represented as an undirected graph $\mathcal{G}$. At each time step, an agent is assumed to either move from one vertex to the other or wait at the current vertex. The duration of both actions is uniform and equals $1$ time step. The plan for the $i$-th agent, $pl_i$, is a sequence of moves, s.t., each move starts where the previous one ends. Two distinct plans have a vertex (or edge) conflict if, at any time step, the agents occupy the same vertex (or traverse the same edge in opposite directions) at that time.

The task is to find a set of $n$ plans, $Pl=\{pl_1, ..., pl_n\}$, s.t. each $pl_i$ starts at $s_i$, ends at $g_i$ and each pair of plans in $Pl$ is conflict-free. The objective to be minimized is typically defined as $SoC(Pl)=\sum_{i=1}^n cost(pl_i)$ (called the \textit{Sum-Of-Costs}) or as $MS(Pl)=\max_{i=1,...,n} cost(pl_i)$ (called the \textit{Makespan}), where $cost(pl_i)$ is the cost of the individual plan which equals the time step when the agent reaches its goal vertex (and does not move away further on).

Notably, two assumptions on how agents behave when they reach their goals are common in MAPF: stay-at-target and disappear-at-target. In the latter case, the agent is assumed to disappear upon reaching its target and, thus, is not able to cause any further conflicts. In this work, we study MAPF under the first assumption (which is intuitively more restrictive). 

\paragraph{MAPF as a sequential decision-making problem} Despite MAPF being typically considered to be a planning problem as defined above, it can also be considered as a sequential decision-making (SDM) problem. Within the SDM framework, the problem is to construct \emph{a policy}, $\pi$, that is a function that maps the current state (the current positions of all agents in the graph) to a (joint) action $\mathbf{a}=a_i \times ... \times a_n$, where $a_i \in A_i$ and $A_i$ is the set of possible actions for agent $i$. When $\pi$ is obtained, it is invoked sequentially until either all agents reach their goals or the threshold on the number of time steps, $t_{max}$, is reached.

For better scalability, the decision-making policy might be decentralized, i.e., each agent chooses its action independently of the other agents. In practice, decentralized agents typically don't have access to the global state of the environment, i.e., positions of the other agents, but rather rely on local observation, $o_t$. For example, if the underlying graph is a 4-connected grid, then the local observation may be a $(2r+1) \times (2r+1)$ patch of the grid centered at the agent's current position (where $r$ is the observation radius) and the latter observes only the agents that are within this patch. A sequence of individual observations and actions forms the agent's history: $h_t=\{o_1, a_1, o_2, a_2, ..., o_{t-1}, a_{t-1}, o_t\}$, where $o_k$ and  $a_k$ denote the action and the observation at time step $k$. This history is typically used to reconstruct a Markovian state of the environment via some approximator $f$: $s_t\approx f(h_t)$ (e.g. $f$ can be represented as a neural network). 

Overall, in the decentralized partially observable setting, the problem is to construct $n$ individual policies of the form: 

\begin{equation*}
\pi_i(s_t) \rightarrow \mathbf{P}(A_i)
\end{equation*}
\noindent where $\mathbf{P}(A_i)$ -- is the probability distribution over the actions. The exact action to be executed at the current time step is considered to be sampled from this distribution.

In this work, we follow a common assumption in MAPF that the agents are homogeneous and cooperative. Thus instead of obtaining $n$ distinct individual policies $\pi_i$, we aim to obtain a single individual policy $\pi$ that governs the behavior of each agent.

\paragraph{Imitation learning} To construct (learn) a decision making policy $\pi$, imitation learning relies on an expert policy $\pi^b$, which is used to collect a set of trajectories: $\mathcal D=\{traj\}$, where each trajectory is composed of the observations and actions: $traj=\{o^b_1,a^b_1,\dots,o^b_L,a^b_L\}$. Intuitively, in MAPF context, $\mathcal D$ represents the expert knowledge on how an agent should behave under different circumstances (it can be obtained running a well-established MAPF solver on a range of problem instances).

Denote now by $\pi_\theta$ a target policy parameterized by $\theta$. The problem of obtaining (learning) this policy from the available data is reduced to the following optimization problem for the parameters $\theta$:
\begin{equation*}
    \theta^\star=\argmin_\theta\mathbb E_{traj\sim\mathcal{D}}\sum_{j=0}^{L}\mathcal L(a_j,a^b_j), 
\end{equation*}
\noindent where $a_j\sim \pi_\theta(s_j)$, $a^b_j\sim \pi_\theta(s^b_j)$, and $\mathcal L$ is a loss function, which is selected depending on the action space. In the considered case, when the action space is discrete, cross-entropy is widely used as $\mathcal{L}$.

\section*{Method}

\begin{figure*}
    \centering    \includegraphics[width=1.0\linewidth]{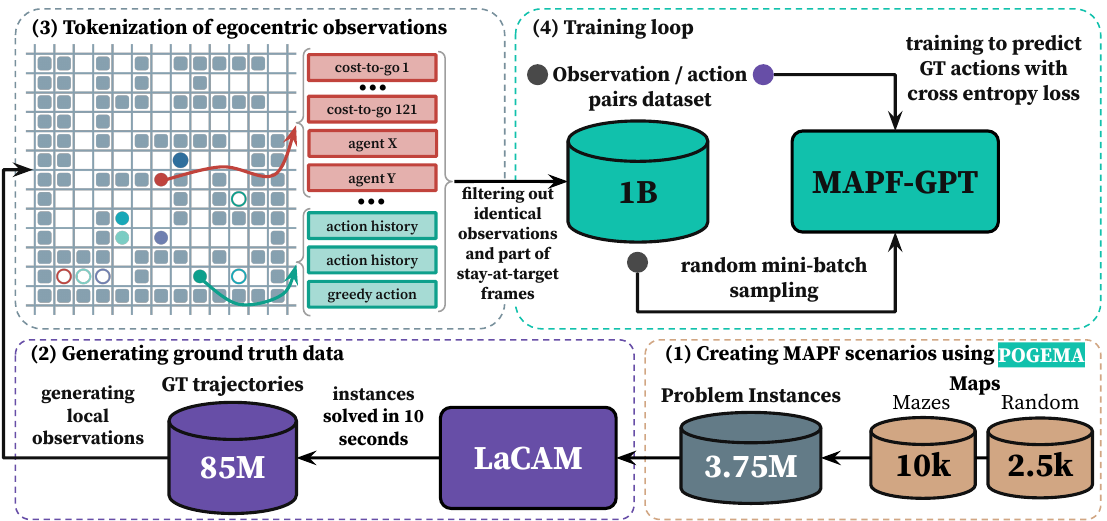}
    \caption{The general pipeline of the MAPF-GPT: (1) Creating MAPF scenarios. (2) Generating ground truth data, i.e. MAPF solutions using an expert solver. (3) Transforming the solutions to the observation-action pairs and tokenization of the observations, which converts them into a format suitable for transformer architectures. (4) Executing the training loop, where observation/action pairs are sampled from the dataset, and the model is trained using cross-entropy loss.}
    \label{fig:scheme}
\end{figure*}

Our approach, MAPF-GPT, is to learn to imitate an expert in solving MAPF. The learning phase of MAPF-GPT consists of the four major steps: creating MAPF scenarios, generating ground truth solutions, tokenizing these solutions, and executing the main training loop -- see Figure~\ref{fig:scheme}. We will now sequentially describe these steps.

\subsection{Creating MAPF Scenarios}

A large, curated dataset is crucial for any data-driven method including ours. To create the set of training instances we used POGEMA~\cite{skrynnik2025pogema}, a versatile tool for developing learnable MAPF solvers that includes utilities to generate maze-like maps and maps with random obstacles, as well as to create MAPF instances from them (i.e. assigning start-goal locations). For our purposes we generate 10K of maze-like maps and 2.5K random maps and further created 3.75M different problem instances on these maps. The size of the maps varies from $17 \times 17$ to $21 \times 21$, the number of the agents is $16$, $24$, or $32$. Please note that as we aim to create an individual policy to solve MAPF in a decentralized fashion (i.e. each agent makes its own decision on how to move based on its local observation) it is not the size of the maps that actually matters but rather the density, i.e., the ratio of the free space to the space occupied by the agents. We use moderate and considerably high densities to make the agents face challenging patterns requiring coordination (especially on the maze-like maps).

\subsection{Generating Ground Truth Data}

To create expert data, we use a recent variant of LaCAM~\cite{okumura2024engineering,okumura2023lacam}, a state-of-the-art MAPF solver that is tailored to quickly find a solution and iteratively enhance it while having a time budget. As we need to solve a large number of MAPF instances (i.e. 3.75M) we set this time budget to be $10$ seconds. 

The output of LaCAM on a single MAPF instance is the set of the individual plans. We further trace each plan and reconstruct an agent's (local) observations in order to form observation-action pairs. We use the following post-processing to filter out some of them. First, if several pairs share the exactly the same observation we keep only one of them (picked randomly). Second, we observe that the fraction of the pairs when an agents waits at the goal is very high, because in many cases numerous agents wait for long times for other agents to reach their goals and just stand still. In fact, almost $40\%$ of the actions in the original expert data are the waiting ones. To remove this imbalance we discard $80\%$ of wait-at-target actions. We end up with 900M  observation-action pairs from the maze-like maps, and 100M -- from the random ones. A 9:1 proportion is chosen due to the maze maps possessing more challenging layouts with numerous narrow passages that require a high degree of cooperation between the agents.

We believe that the obtained dataset composed of 1B observation-action pairs is currently the largest dataset of such kind and may bring value to the other researchers developing learnable MAPF solvers.  Additional technical details are presented in the Appendix~\ref{appendix:dataset} of the paper.

\subsection{Tokenization} 

\begin{figure}[htb!]
    \centering
    \includegraphics[width=0.9\linewidth]{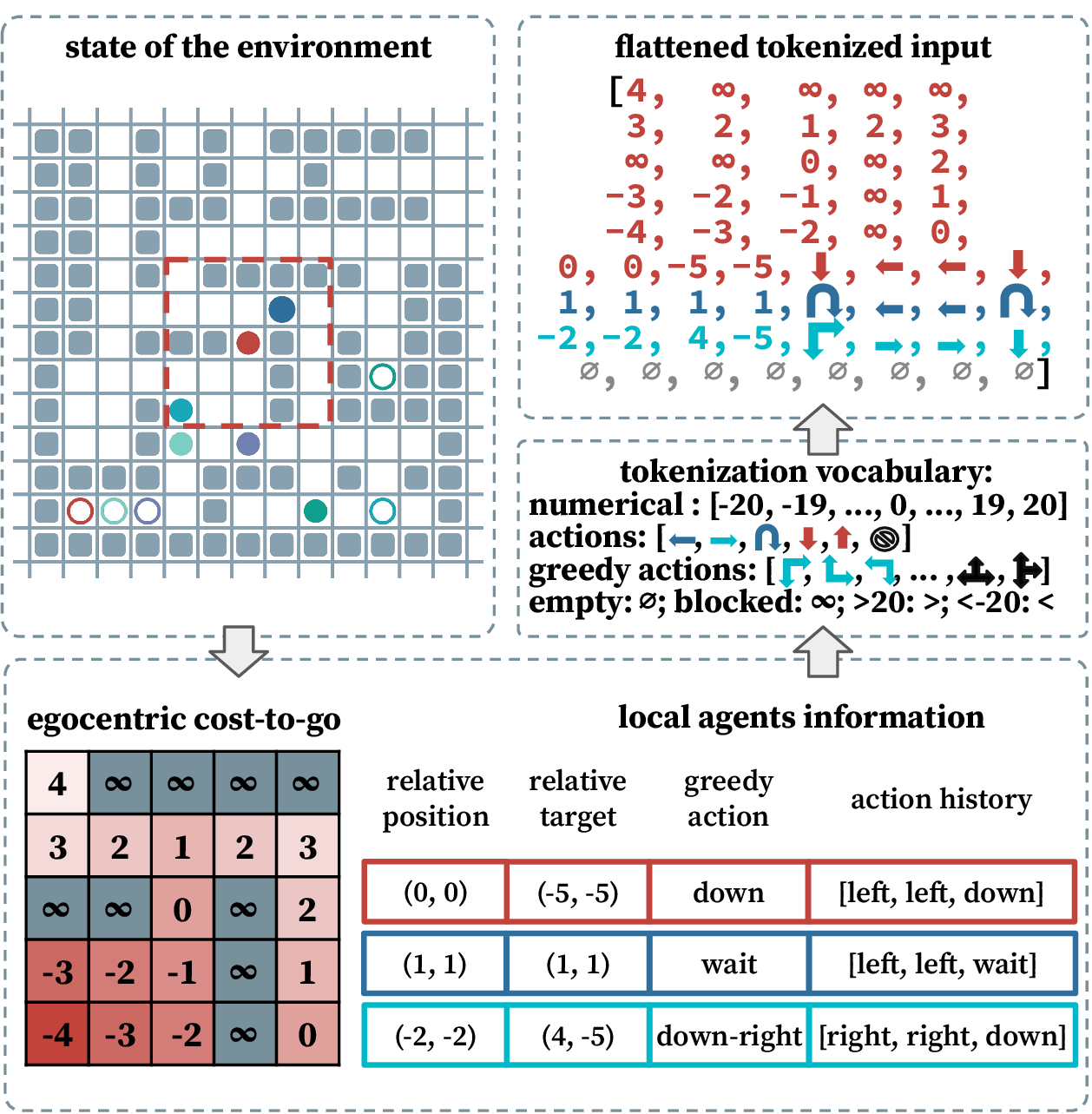}
    \caption{
    The tokenization process for the MAPF-GPT model uses a vocabulary of 67 tokens, with an input of 256 tokens. Fewer tokens are shown for clarity and visibility.
    }
    \label{fig:tokenization}
    \vspace{-10px}
\end{figure}

Tokenization can be thought of as the process of transforming the data, observation-action pairs in our case, into the sequence of special symbols, tokens, to be further fed to the neural network that is trained to predict a single token, i.e. action, from the sequence of the input tokens, i.e. the ones that encode the observation. The input tokens are typically referred to as the \emph{context}. We now wish to describe how our context, i.e. the observation, is structured.

The local observation of an agent at a certain time step while following its (expert) path is composed of two parts. The first one relates to the map in the vicinity of the agent, i.e. which parts are traversable, which are not, and going to which areas moves the agent closer to its goal. As we used grids to represent the environment the local field-of-view is composed of a square patch of the cells centered at the agent's current position. For each traversable cell we compute its cost-to-go value, i.e. the cost of the shortest path to the cell from the goal location. As the cost of this path might be arbitrarily, we normalize it. I.e. the cost-to-go value is set to $0$ for the cell the agent is currently in, $(x_{cur}, y_{cur})$. The values for the other traversable cells within the field-of-view are computed as $\text{cost-to-go}(x, y) - \text{cost-to-go}(x_{cur}, y_{cur})$, where the coordinates are absolute w.r.t. the global coordinate frame. The blocked cells are assigned infinite values. 

The second part of the observation contains data about the agent itself and the nearby ones. The information about each agent consists of the coordinates of its current and goal locations, actions history, i.e., the actions that were made in the previous $k$ steps, and an action that is preferable w.r.t. the agent's individual cost-to-go map -- the so-called greedy action. Please note that there may be cases where more than one action leads to a decrease in cost-to-go. Thus, we use special markers to indicate these multi-direction greedy actions (e.g., ``up-right'').

The input of the model consists of $256$ tokens that encode the local observation of the agent. For the first part, i.e., cost-to-go values, we use the $11 \times 11$ field of view, which results in $121$ tokens.  An example of the tokenization mechanism is illustrated in Figure~\ref{fig:tokenization}.

The rest of the input ($135$ tokens) is devoted to the information about agents. As it's important to consider only the agents that can potentially influence the egocentric agent, we consider only the ones that are located in the $11 \times 11$ field of view at the current time step. The information about each agent is encoded via $10$ tokens: $2$ for the current position, $2$ for goal location, $5$ for action history, and $1$ for the next greedy action. Thus, we are able to encode the information about at most $13$ agents, including the egocentric one. The rest of $5$ tokens in the input are encoded with the empty token. In case there are not enough agents in the local field of view, the information about missing agents is also filled with empty tokens. The information about agents is sorted based on the distance to the egocentric agent, i.e., the information about the egocentric agent itself always goes first, as its distance is always $0$.

Information in the observation includes both numerical values, such as cost-to-go values or coordinates, and some literal ones that, for example, correspond to the actions. We have chosen the range $[-20, 20]$ for the numerical values, i.e., $41$ different tokens. This range was chosen due to the size of the maps used in the training dataset, which is at most $21 \times 21$. However, the cost-to-go values might go beyond this range. For this purpose, we utilize $2$ additional tokens for the values that are beyond $20$ or below $-20$. The coordinates also might be clipped if their values go outside this range. There is also a token that corresponds to the $\infty$ value for blocked cells. The agents are allowed to perform $5$ actions -- to move into $4$ directions and to wait in place. We have also added one additional token to encode the empty action, i.e., when there are not enough actions performed at the beginning of the episode. The information about the next greedy action cannot be directly encoded with a single token utilizing the tokens that represent the actions due to the fact that there might be two or more directions that reduce the cost-to-go value. To cover all possible cases, such as up-right, left-down-right, etc., we have added $16$ more tokens. The last one is an empty token used for padding to $256$ tokens. In total the vocabulary consists of $67$ different tokens.

\begin{figure*}[thb!]
    \centering
    \includegraphics[width=1.0\linewidth]{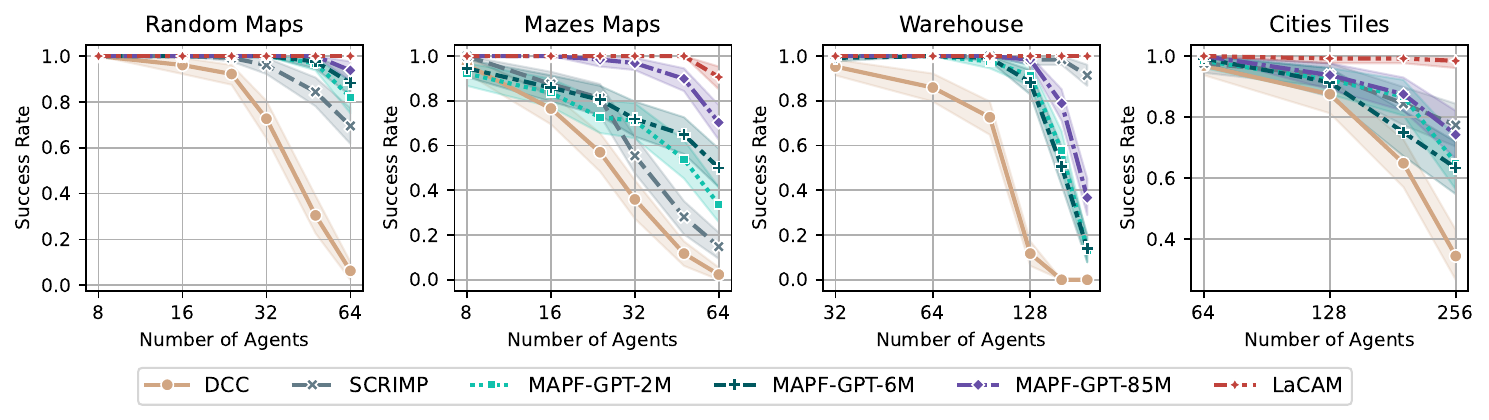}    
    \caption{Success rate of the evaluated MAPF solvers on different maps. The shaded area indicates $95\%$ confidence intervals.}
    \label{fig:experiments-success-rate}
\end{figure*}

\begin{figure*}[thb!]
    \centering
    \includegraphics[width=1.0\linewidth]{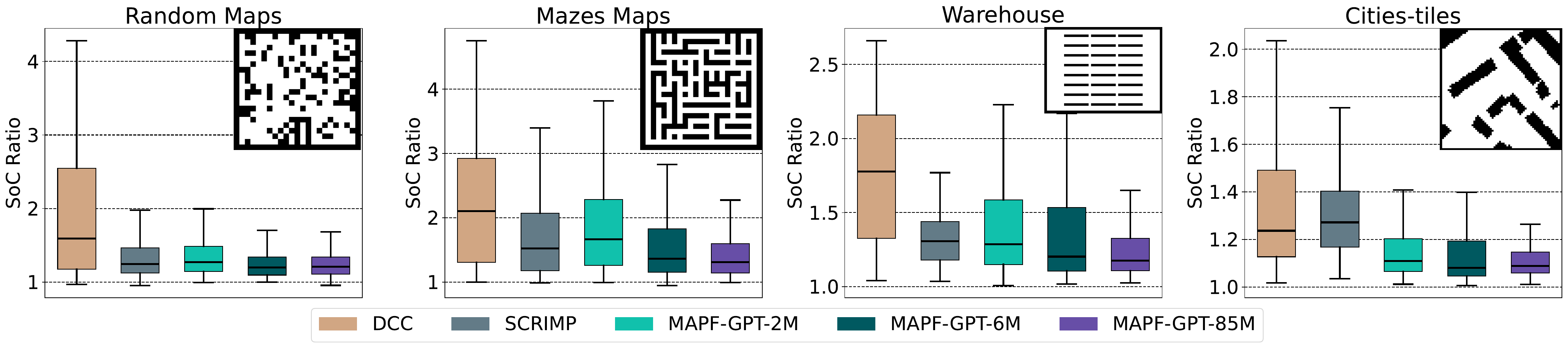}    
    \caption{Quality of the obtained solutions relative to the ones of LaCAM (lower is better). }
    \label{fig:relative-to-lacam}
\end{figure*}
%

\subsection{Model Training}

As the model backbone, we used a modern decoder-only transformer~\cite{brown2020language}. We used a softmax layer to parameterize the discrete probability distribution. To sample an action, we then used multinomial sampling. The length of the input sequence (context size) of the model is $256$. The output size of the model is $5$ since each agent has $5$ available discrete actions. We used learnable position embeddings. We don't use causal masking, which is common practice in the NLP~\cite{radford2019language}, since the model predicts only a single action ahead in a non-autoregressive manner. To speed up training, we used the flash attention technique~\cite{dao2024flashattention}.

We use models of different sizes (i.e. number of parameters) in our work. Specifically, our largest model contains $85$M parameters. We also consider much smaller models that contain $6$M and $2$M parameters.

\paragraph{Training protocol} The model was trained to replicate the behavior of the expert policy using cross-entropy loss (i.e., log-loss) via mini-batch stochastic gradient descent, optimized with AdamW~\cite{loshchilov2018decoupled}. The target label for this loss is a ground-truth action index provided by the expert policy, LaCAM.  LaCAM is a centralized solver that builds a path for all agents during the whole episode, leveraging information about the full environment state. In contrast, the trainable model relies solely on a local observation $o$ of each agent $u$.

\begin{equation}
    -\log \textbf{p}_{\theta}\left(a_u^{LaCAM}(s) \mid o_u\right).
\end{equation}

Once trained, this policy enables the sampling of actions from it. While an alternative could be to pick the action with the highest probability, we use sampling, recognizing the decentralized nature of the policy. Sampling selects actions based on the probability distribution given by the policy:

\begin{equation}
    \hat{a}^u(o_u) \sim \textbf{p}_{\theta}(o_u)
\end{equation}

\noindent where \(\textbf{p}_{\theta}(o_u)\) represents the probability distribution over actions computed by the model for the observation $o_u$.

We used $2000$ warm-up iterations and cosine annealing~\cite{loshchilov2017sgdr}, with a gradient clipping value of $1.0$ and a weight decay parameter of $0.1$. The entire 1B dataset was used to train the 85M model, which underwent 1M iterations with a batch size of $512$, resulting in $15.625$ epochs based on the gradient accumulation steps, set at $16$. For training the 6M and 2M models, we used portions of the 1B dataset -- 150M and 40M, respectively. Additional details about the parameters influencing the training process are provided in the Appendix~\ref{appendix:hyperparameters}.

\section*{Experimental Evaluation} 

\paragraph{Main results} In the first series of experiments, we compare 3 variants of MAPF-GPT varying in the number of parameters in their neural networks (2M, 6M, 85M) with the state-of-the-art learnable MAPF solvers: DCC~\cite{ma2021learning} and SCRIMP~\cite{wang2023scrimp}\footnote{Please note that while a plethora of learnable MAPF solvers exist, only a few are tailored to the studied setting: MAPF with non-disappearing agents.}. We use the pre-trained weights for DCC and SCRIMP. These weights were obtained by the authors while training on the random maps. 
Additionally, we present the results of LaCAM~\cite{okumura2024engineering}, which served as the expert centralized solver for data collection. For evaluation we used \texttt{Random}, \texttt{Mazes}, \texttt{Warehouse}, \texttt{Cities-tiles} maps (the latter two are out-of-distribution for all learnable solvers). All maps and instances utilized during the evaluation were taken from \cite{skrynnik2025pogema}. The details on the maps and problem instances are given in the Appendix~\ref{appendix:benchmark}.

The results are presented in Figure~\ref{fig:experiments-success-rate}, where the success rate of all solvers is shown\footnote{Please note that the previous arXiv version of the paper, as well as the version accepted to AAAI'25, contains different results for SCRIMP. This is due to a technical error occurred when running SCRIMP on our test instances. Unfortunately, the error was identified by us after the AAAI'25 conference.}. Clearly, all variants of MAPF-GPT outperform both DCC and SCRIMP on \texttt{Random} and \texttt{Mazes} maps. On \texttt{Cities-tiles} the success rate of MAPF-GPT-85M is better than the one of DCC and is on par with SCRIMP. On \texttt{Warehouse} the latter solver is superior to all others when the number of agents exceeds 128. A possible explanation is that SCRIMP utilizes the so-called value-based tie-breaking mechanism, allowing the agents to iteratively re-select the actions, and this mechanism turns out to pe particularly valuable to the \texttt{Warehouse} setup.   


Figure~\ref{fig:relative-to-lacam} shows the Sum-of-Costs (SoC) achieved by the solvers relative to the SoC of LaCAM (the lower - the better). As can be seen, MAPF-GPTs outperform the other approaches and their performance correlates with the number of model parameters. Interestingly, there are some rare cases on \texttt{Random} maps where DCC and MAPF-GPT-85M outperformed LaCAM in terms of SoC. The same situation is observed for MAPF-GPT-6M on \texttt{Mazes} maps.

\paragraph{Ablation study} In this experiment, we study how each part of the information influences the performance of the MAPF-GPT agent. To address this, we trained a 6M parameter model with certain pieces of information masked that were provided to the original model. We examine four different cases: if there is no goal information for all agents (noGoal), if there is no greedy action provided (noGA), if there is no action history (noAH), and if the agent is trained without cost-to-go information (still retaining information about obstacles). We used an additional type of map, \texttt{Puzzles}, for this experiment. 
These maps are quite small ($5\times5$) and are specifically designed to assess the capability of algorithms in solving complex scenarios where agents need to execute cooperative actions. The results are presented in Table~\ref{tab:ablation}.

\begin{table}[ht!]
    \centering
    \resizebox{\columnwidth}{!}{
        \begin{tabular}{lccccc}
            \toprule
            Scenario   & 6M              & noGoal          & noGA   & noAH            & noC2G  \\
            \midrule
            \midrule
            Random    & \bc 97.6\% & 95.7\%          & 97.0\% & 95.6\%          & 25.8\% \\
            Mazes     & 74.6\%          & 71.6\%          & 37.6\% & \bc 85.8\% & 15.1\% \\
            Warehouse & 94.1\%          & 92.8\%          & 87.7\% & \bc 94.8\% & 11.5\% \\
            Cities-tiles  & 82.0\%          & \bc 88.4\% & 79.1\% & 82.2\%          & 10.2\% \\
            Puzzles   & \bc 94.0\% & 92.7\%          & 92.7\% & 91.5\%          & 52.5\% \\
            \bottomrule
        \end{tabular}
    }
    \caption{Success rates of different versions of MAPF-GPT-6M on all sets of maps from POGEMA benchmark.}
    \label{tab:ablation}
\end{table}

As can be seen, the 6M model without masking shows better results on the \texttt{Random} and \texttt{Puzzles} maps. The model trained without goal information shows better results on the \texttt{Cities-tiles} maps, with $88.4\%$ compared to $82.0\%$ for the original model. This improvement is likely due to the large map sizes, where most conflicts arise during the agents' movement toward their goals, making the exact goal coordinates less critical. Additionally, the functionality was compensated by the greedy action, which indicates the direction to the goal.

Surprisingly, the model trained without action history shows better performance on the \texttt{Mazes} and \texttt{Warehouse} instances. This suggests that action history is not crucial for behavioral cloning, as LaCAM does not rely on it. Despite these results, we argue for retaining movement history information in the agent's observation, which could be essential for further fine-tuning (we leave this for future work).

Masking greedy action and cost-to-go information degrades the performance of the model on all testing tasks, highlighting their crucial role in effective pathfinding and conflict resolution.

\paragraph{LifeLong MAPF} In addition to evaluating MAPF-GPT on the MAPF instances it was trained on, we also assessed its performance in the Life-Long MAPF (LMAPF) setup. Unlike regular MAPF problems, in LMAPF, each agent receives a new goal location every time it reaches its current one. In this setup, the primary objective is throughput, which is defined as the average number of goals reached by all agents per time step.

\begin{table}[b!]
    \centering
    \resizebox{\columnwidth}{!}{
    \begin{tabular}{lccccc}
        \toprule
        Scenario  & 6M    & 6M tuned & RHCR      & Follower  & MATS-LP \\
        \midrule
        \midrule
        Random    & 1.497 & 1.507    & \bc 2.164 & 1.637     & 1.674   \\
        Mazes     & 0.908 & 1.087    & \bc 1.554 & 1.140     & 1.125   \\
        Warehouse & 1.113 & 1.270    & 2.352     & \bc 2.731 & 1.701   \\
        Cities-tiles  & 2.840 & 2.994    & \bc 3.480 & 3.271     & 3.320   \\
        \bottomrule
    \end{tabular}
    }
    \caption{Average throughput (higher is better) of MAPF-GPT-6M both in zero-shot mode and after fine-tuning compared to RHCR, Follower and MATS-LP.}
    \label{tab:lmapf}
\end{table}

We evaluated MAPF-GPT in both zero-shot and fine-tuned configurations. To fine-tune the model, we generated an additional dataset using \texttt{Mazes} maps. For expert data, we employed the RHCR approach~\cite{li2021lifelong}, as LaCAM is not well-suited for LMAPF. The dataset contains 90 million observation-action pairs. We used MAPF-GPT-6M for this experiment.

The results are presented in Table~\ref{tab:lmapf}. Even the zero-shot model is able to compete with other existing learning-based approaches, such as Follower~\cite{skrynnik2023learn} and MATS-LP~\cite{skrynnik2024decentralized}. Moreover, in all cases, fine-tuning improved the results of MAPF-GPT-6M.

These results demonstrate the ability of MAPF-GPT to perform \emph{zero-shot} learning, i.e., the ability to solve types of problems that it was not initially designed for (e.g., solving LMAPF instead of MAPF), and that \emph{fine-tuning} MAPF-GPT is indeed possible, i.e., additional training on new types of tasks increases its performance in solving these tasks.

\begin{figure}[htb!]
    \centering
    \includegraphics[width=1.0\linewidth]{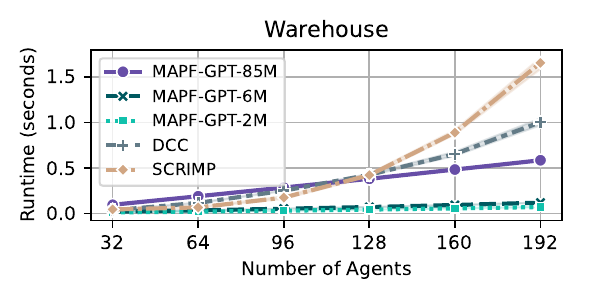}
    \caption{Runtime of MAPF-GPT, DCC, and SCRIMP models on the \texttt{Warehouse} map. The plot shows the average time required to decide the next action for all agents as the number of agents increases. }
    \label{fig:runtime}
\end{figure}

\paragraph{Runtime} In this experiment, we compare the runtime of the considered solvers. The results are presented in Figure~\ref{fig:runtime}; each data point indicates the average time spent deciding the next action for all agents. All MAPF-GPT models scale linearly with the increasing number of agents. The largest model, MAPF-GPT-85M, shows a slightly higher runtime than DCC and SCRIMP on the instances with up to 96 agents. However, beyond 128 agents, the runtime of MAPG-GPT-85M is better, as it depends linearly on the number of agents. Notably, the MAPF-GPT-2M and MAPF-GPT-6M models are more than 13 times faster than SCRIMP and 8 times faster than DCC for 192 agents setup.

\section*{Conclusion}

In this work, we studied the MAPF problem as a sequential decision-making task. We proposed the approach to derive an individual policy based on state-of-the-art machine learning techniques, specifically (supervised) imitation learning from expert data -- MAPF-GPT. To train MAPF-GPT, we created the comprehensive dataset of expert MAPF solutions, transformed these solutions into observation-action pairs, tokenized them, and trained several transformer models (with varying numbers of parameters) on this data. Empirically, we demonstrate that even with a quite moderate number of parameters, such as 2M, MAPF-GPT significantly outperforms modern learnable MAPF competitors across the wide range of setups. Our results provide a clear positive answer to the question, ``Is it possible to create a strong learnable MAPF solver purely through imitation learning?''. The limitations of our approach are discussed in the Appendix~\ref{appendix:limitations}.

\newpage
\bibliography{aaai25}
\appendix

\section*{Appendix}

\section{Implementation Details}
\label{appendix:implementation_details}

The open-sourced repository contains all the necessary resources to replicate the results of \textsc{MAPF-GPT}. This includes scripts for dataset generation, training, evaluation. The codebase also provides example scripts to help users quickly understand and run \textsc{MAPF-GPT}, along with the detailed instructions for setting up the environment using Docker. Additionally, configuration files for benchmarks and datasets are included to facilitate effortless replication of the experiments.

The GPT model’s code is primarily based on the NanoGPT codebase\footnote{https://github.com/karpathy/nanoGPT}. NanoGPT was chosen for its simple yet modular design, making it easier to modify and adapt the code, including potential adjustments for fine-tuning in environment (e.g., with reinforcement learning).

\section{Hyperparameters and Training Details}
\label{appendix:hyperparameters}

This section details the hyperparameters used while training all three our models: MAPF-GPT-2M, MAPF-GPT-6M, and MAPF-GPT-85M. Common parameters for the models are listed in Table~\ref{tab:common_hyperparameters}. 

\begin{table}[htbp]
\centering

\begin{tabular}{lc}
\toprule
\textbf{Parameter} & \textbf{Value} \\
\midrule
\midrule
Minimum learning rate  & 6e-5 \\
Maximum learning rate  & 6e-4 \\
Learning rate decay  & cosine  \\
Warm-up iterations  & 2000 \\
AdamW optimizer beta1  & 0.9 \\
AdamW optimizer beta2  & 0.95 \\
Gradient clipping  & 1.0 \\
Weight decay  & 1e-1 \\
Data type for computations & float16 \\
Use PyTorch 2.0 compilation & True \\
Gradient accumulation steps & 16 \\
Block size & 256 \\
\bottomrule
\end{tabular}
\caption{Common hyperparameters for MAPF-GPT models.}
\label{tab:common_hyperparameters}
\end{table}

Table~\ref{tab:model_specific_hyperparameters} provides model-specific hyperparameters for MAPF-GPT-2M, MAPF-GPT-6M, and MAPF-GPT-85M. We didn't change the hyperparameters between the models except for those that influence their size and the duration of the learning process.
Training the 85M model for 1M iterations took 243 hours using 4x H100 80GB NVIDIA GPUs. Training the 6M model for 30K iterations took 50 hours using 2x A100 80GB NVIDIA GPUs. The 2M model was trained on a single H100 80GB NVIDIA GPU within 12 hours. It is also worth noting that while training the 6M and 2M models we could not fully utilize the computational power of the corresponding GPUs, as the bottleneck was actually in data processing and transferring it to the GPU.

\begin{table}[htbp]
\centering
\resizebox{\columnwidth}{!}{
\begin{tabular}{lccc}
\toprule
\textbf{Parameter} & \textbf{2M} & \textbf{6M} & \textbf{85M} \\
\midrule
\midrule
Number of epochs & 46.875 & 12.5 & 15.625 \\
Total training iterations  & 15,000 & 30,000 & 1,000,000 \\
Batch size  & 4096 & 2048 & 512 \\
Number of layers  & 5 & 8 & 12 \\
Number of attention heads & 5 & 8 & 12 \\
Embedding size  & 160 & 256 & 768 \\
\bottomrule
\end{tabular}
}
\caption{Model-specific hyperparameters.}
\label{tab:model_specific_hyperparameters}
\end{table}

\begin{figure}[ht!]
    \centering
    \includegraphics[width=0.85\linewidth]{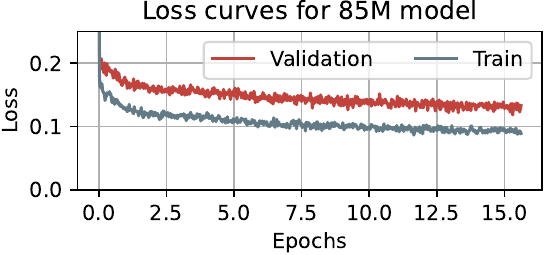}
    \caption{Training and validation loss curves for MAPF-GPT-85M.}
    \label{fig:loss-curves}
\end{figure}

The loss curves for MAPF-GPT-85M are presented in Figure~\ref{fig:loss-curves}, showing the averaged loss for both training and validation. The validation dataset contains an equal 1:1 ratio of (observation, action) pairs between maze-like and random maps, in contrast to the 9:1 ratio in the training dataset.

\begin{figure}[ht!]
    \centering
    \includegraphics[width=0.495\linewidth]{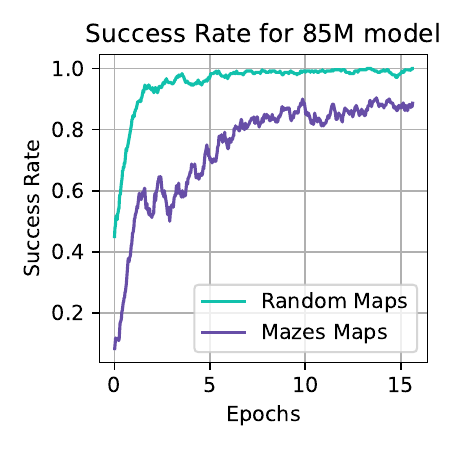}
    \includegraphics[width=0.495\linewidth]{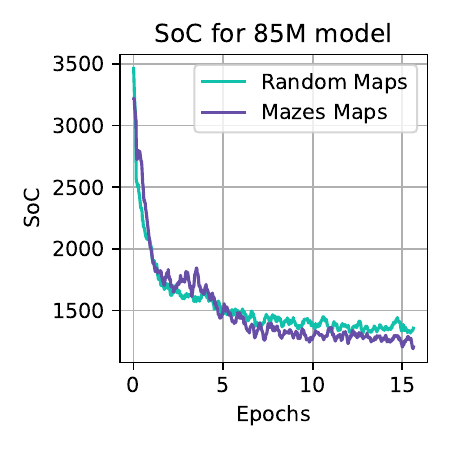}
    
    \caption{Aggregated evaluation results for Success rate  and Sum of Costs during training.}
    \label{fig:train-csr-soc}
\end{figure}

During training, we also performed evaluations in the environment using intermediate checkpoints. The aggregated results of these evaluations are presented in Figure~\ref{fig:train-csr-soc}. Intermediate evaluations were conducted on 6 random maps with 48 and 64 agents, as well as 6 maze maps with 32 and 48 agents. These maps were taken from the set used to generate the validation dataset.

\section{Dataset: Technical Details}
\label{appendix:dataset}

As mentioned in the main part of the text, to generate the 1B training dataset, we first generated 10,000 maps using a maze generator and 2,500 maps using a random map generator. Next, we generated instances with 16, 24, or 32 agents, with 100 seeds (start-goal locations) per map, resulting in 3.75M different instances.

In the next step, we ran the centralized MAPF solver, i.e. LaCAM, to obtain the expert data. We used the latest version of LaCAM and its C++ implementation provided by the authors\footnote{https://github.com/Kei18/lacam3}. LaCAM was given 10 seconds to solve each instance in a single-thread mode. We ran LaCAM in parallel on two workstations with AMD Ryzen Threadripper 3970X 32-Core Processors, which took nearly 100 hours to obtain the data. Around 3\% of the instances were not successfully solved (primarily those with 32 agents) due to the presence of unsolvable tasks and a time limit.

The resulting log files, which contain the actions taken by the agents at each time step, were evenly distributed into $50$ chunks. The logs from each chunk were processed to generate local observations for every agent. At this stage, we removed redundant waiting actions and duplicate observations as detailed in the main body of the paper.

In the final step, the obtained data was shuffled and saved into $10$ .arrow files per chunk. Each .arrow file contains $2^{21}$(slightly more than 2 million) (observation, action) pairs, with 10\% of the data coming from the random maps and 90\% -- from mazes.

The resultant 1B dataset contains 500 such files, requiring 258GB of disk space. While training the MAPF-GPT-6M and MAPF-GPT-2M models, we randomly selected 75 and 20 files from this dataset, resulting in 150M and 40M datasets, respectively. The full dataset was used for training MAPF-GPT-85M model.

\section{Benchmark}
\label{appendix:benchmark}

For empirical evaluation we have utilized the evaluation tools provided by the POGEMA benchmark~\cite{skrynnik2025pogema}. Specifically, we used the following set of maps: \texttt{Random}, \texttt{Mazes}, \texttt{Warehouse}~\cite{li2021lifelong}, \texttt{Cities-tiles}~\cite{stern2019multi} and \texttt{Puzzles}. Table \ref{tab:exp_details} contains information about the number of agents, maps, and instances per map (seeds) used in each set.

\begin{table}[ht!]
    \centering
    \resizebox{\columnwidth}{!}{
    \begin{tabular}{rccccc}
    \toprule
    Set &  Agents &  Maps & Map Size & Seeds &  Steps \\
    \midrule
    \midrule
      \texttt{Random} & 8, 16, 24, 32, 48, 64 & 128 & 17$\times$17 - 21$\times$21& 1 & 128 \\
      \texttt{Mazes} & 8, 16, 24, 32, 48, 64 & 128 & 17$\times$17 - 21$\times$21& 1 & 128 \\
      \texttt{Warehouse} & 32, 64, 96, 128, 160, 192 & 1 & 33$\times$46& 128 & 128 \\
      \texttt{Cities-tiles}& 64, 128, 192, 256 & 128& 64$\times$64 & 1& 256 \\
      \texttt{Puzzles} & 2, 3, 4 & 16 & 5$\times$5& 10 & 128\\
    \bottomrule
    \end{tabular}
    }
    \caption{Details about different sets of maps from POGEMA benchmark.}
    \label{tab:exp_details}

\end{table}

\begin{figure*}[htb!]
    \centering
        \begin{subfigure}[b]{0.28\textwidth}
        \centering
        \includegraphics[height=140px]{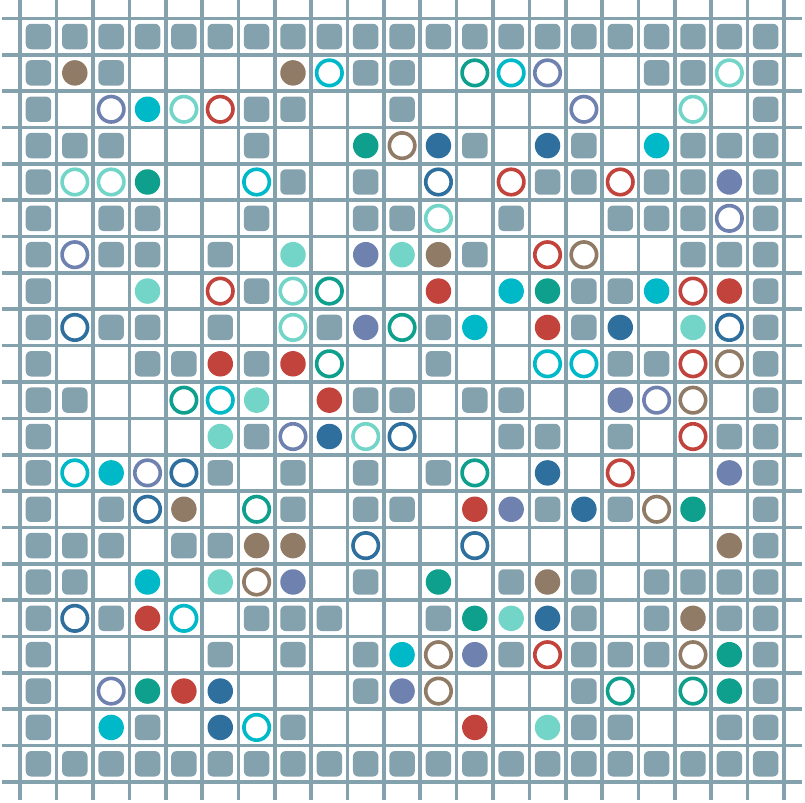}
        \caption{\texttt{Random}}
        \label{fig:appendix:random}
    \end{subfigure}
    \hspace{10px}
    \begin{subfigure}[b]{0.28\textwidth}
        \centering
        \includegraphics[height=140px]{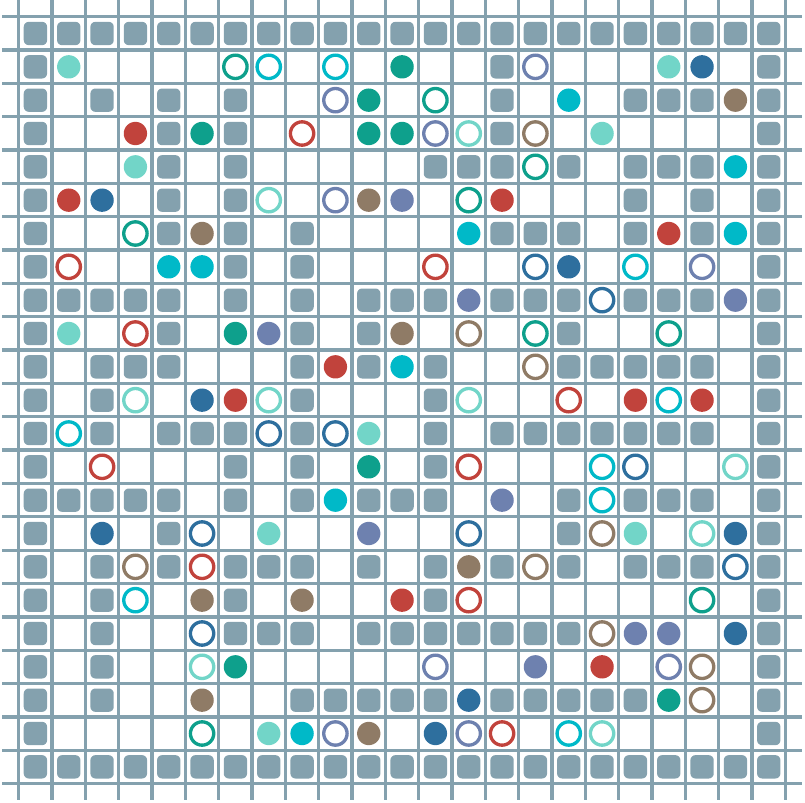}
        \caption{\texttt{Mazes}}
        \label{fig:appendix:maze}
    \end{subfigure}
    \hspace{10px}
    \begin{subfigure}[b]{0.38\textwidth}
        \centering
        \includegraphics[height=140px]{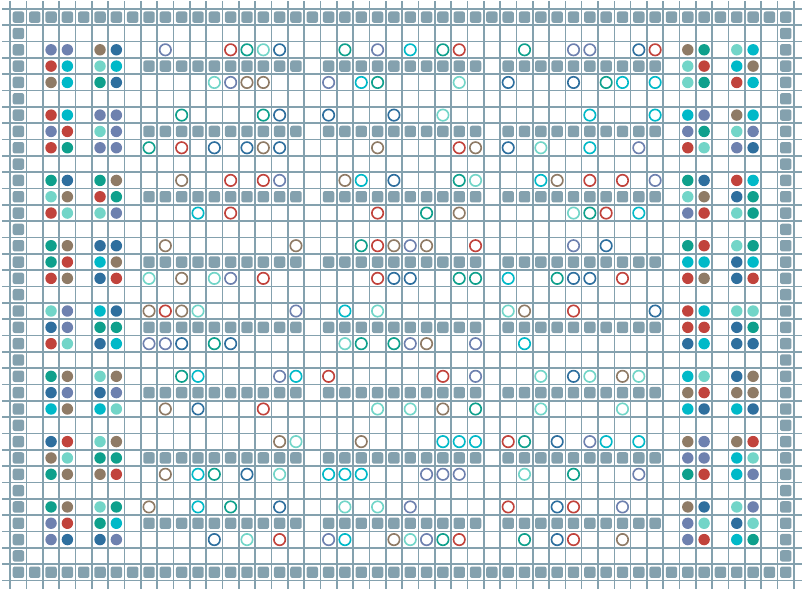}
        \caption{\texttt{Warehouse}}
        \label{fig:appendix:puzzle}
    \end{subfigure}

    \caption{The examples of maps from the POGEMA benchmark. Plots show the maximum number of agents tested on each map: 64 agents for \texttt{Random} and \texttt{Mazes} sets, and 192 agents for \texttt{Warehouse} scenario.}
    \label{fig:appendix:maps_examples}
\end{figure*}

Figure~\ref{fig:appendix:maps_examples} showcases examples of maps from the POGEMA benchmark. The \texttt{Mazes} and \texttt{Random} map sets are generated using the built-in generators within POGEMA. The \texttt{Puzzles} maps are handcrafted. The \texttt{Warehouse} is a single map with limitations on possible start and goal positions, which has been used in previous LMAPF-related studies~\cite{li2021lifelong, skrynnik2023learn}. Start locations on this map can only be generated on the left or right sides of the map, where there are no obstacles, while goal locations can only be placed beyond or below the obstacles in the center of the map. These constraints prevent the generation of instances with more than 192 agents on this map. 

\begin{figure}[htb!]
    \centering
    \includegraphics[width=1.0\linewidth]{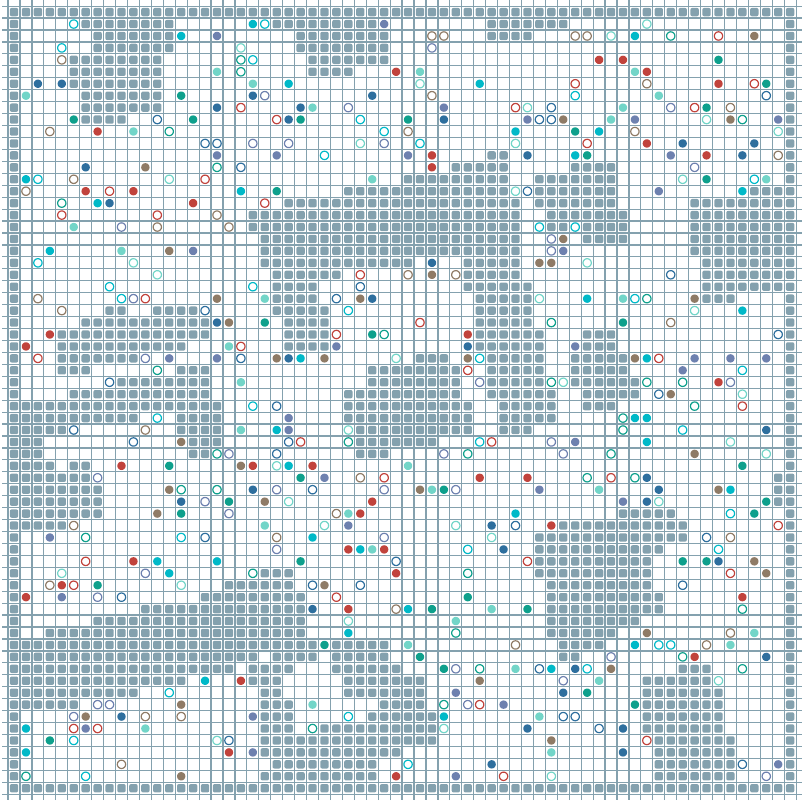}
    \caption{Example of a map from the \texttt{Cities-tiles} set in the POGEMA benchmark. The original city maps, with dimensions of $256 \times 256$, were divided into 16 tiles, resulting in smaller $64 \times 64$ maps.}
    \label{fig:example-moving-ai}
\end{figure}

The final evaluated set of maps is \texttt{Cities-tiles}. This set of maps, as well as the others, is taken from POGEMA benchmark, however, these maps are based on the ones presented in the MovingAI benchmark. While the original MovingAI dataset includes various types of maps, including random and maze maps, the \texttt{Cities-tiles} set in POGEMA contains only city maps. Additionally, due to the large size ($256\times256$) of city maps, they were divided into 16 tiles, resulting in maps of $64\times64$ size. An example of such scenario is presented in Figure~\ref{fig:example-moving-ai}.

The evaluation of all the approaches, including the MAPF-GPT models and the baselines, was performed on the same workstation equipped with AMD Ryzen Threadripper 3970X 32-Core Processors, 256GB of RAM, and 2x RTX 3080 Ti 12GB NVIDIA GPUs.

It's important to note that during the runtime experiments, only one GPU was utilized, and the tasks were run in sequential mode, i.e., no multiple tasks were executed in parallel during this experiment. For this evaluation, we ran all approaches on the \texttt{Warehouse} map with 32 to 192 agents, using five different seeds, and averaged the obtained results.

\section{Limitations} 
\label{appendix:limitations}

The main limitation of MAPF-GPT is a generic one, shared with the other learnable methods, i.e. it lacks theoretical guarantees. The next limitation is that training large models, e.g., MAPF-GPT-85M, is quite demanding (requires expensive hardware and prolonged time). Still, our smaller models containing 6M and 2M parameters are much less demanding while providing competitive results. It should also be noted that all MAPF-GPT models are sensitive to the quality of trajectories in the expert data set. As has been shown in previous research on behavior cloning with transformers, e.g.~\cite{chen2021decision}, adding low-quality trajectories to expert data may lead to significant degradation in model's performance. 
It is also unclear how effectively MAPF-GPT can replicate the behavior of the other existing centralized approaches (such as CBS~\cite{sharon2015conflict} that is an optimal MAPF solver). This dependence on the type of behavioral expert policy requires further research.

\section{Evaluation on Puzzles}
\label{appendix:puzzles}

\begin{figure}[htb!]
    \centering
    \includegraphics[width=0.242\linewidth]{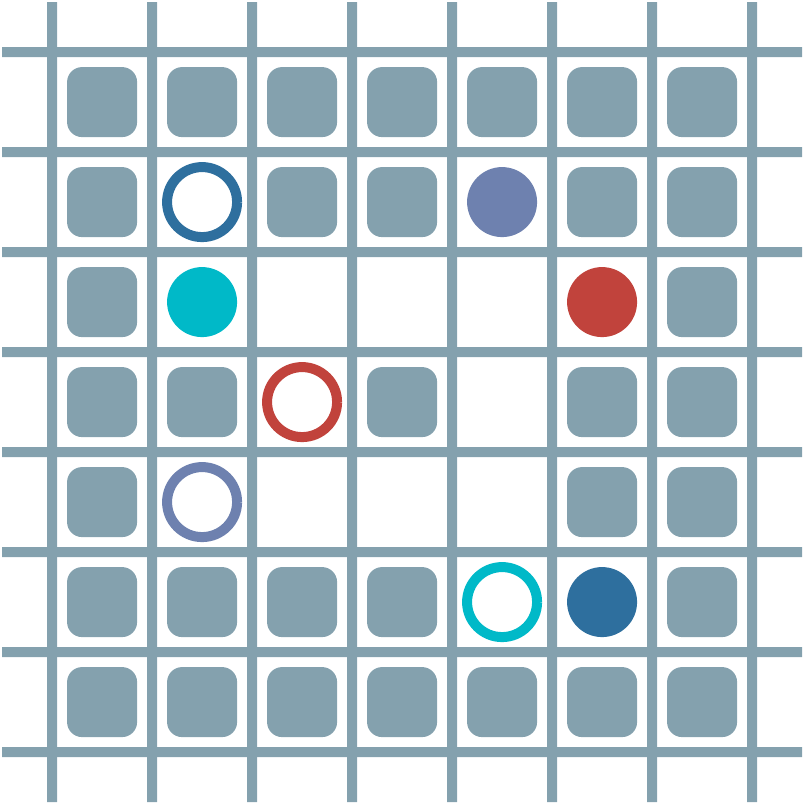}
    \includegraphics[width=0.242\linewidth]{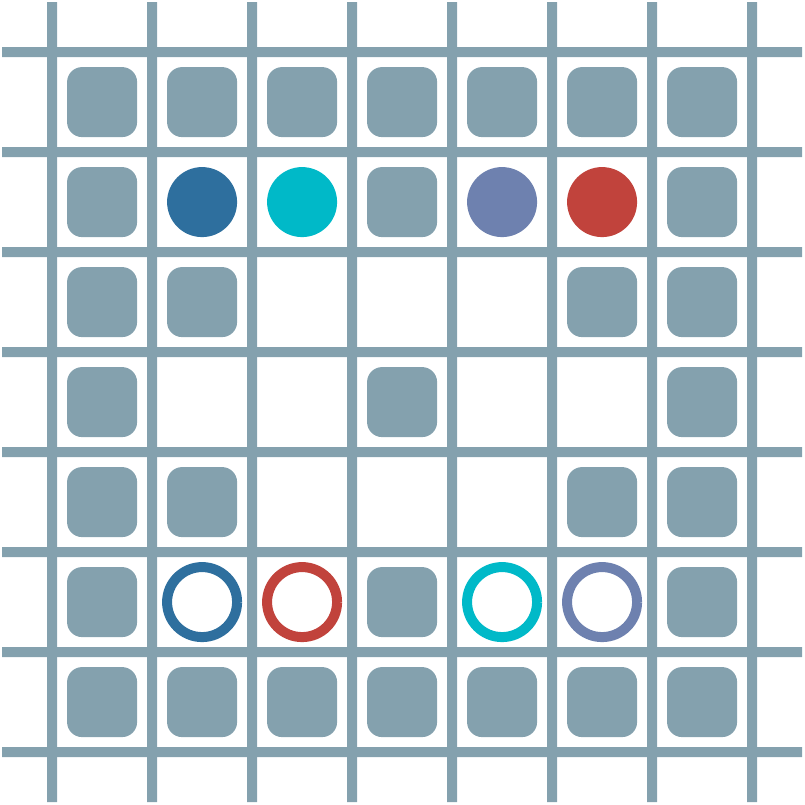}
    \includegraphics[width=0.242\linewidth]{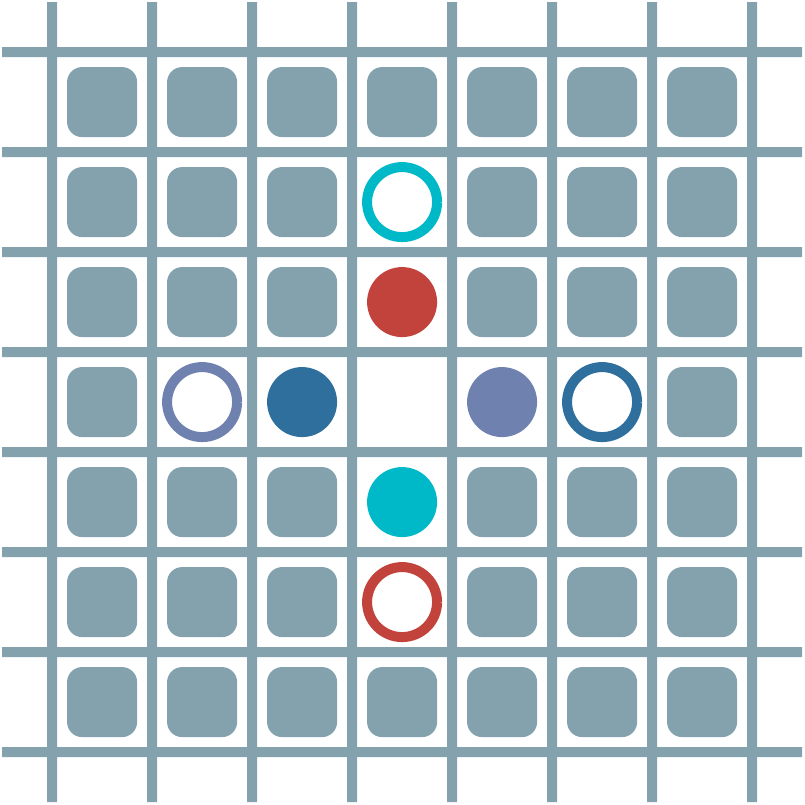}
    \includegraphics[width=0.242\linewidth]{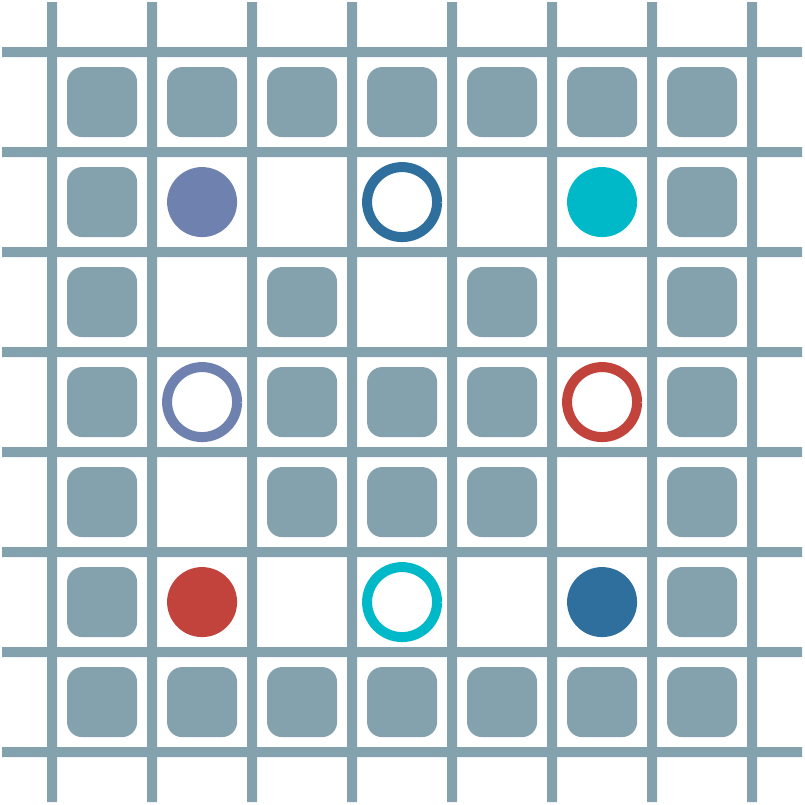}
    
    \caption{Examples of \texttt{Puzzles} map instances. This set of scenarios is particularly challenging and requires high cooperation between agents. }
    \label{fig:puzzles}
\end{figure}

The \texttt{Puzzles} set is part of the POGEMA benchmark, briefly mentioned in the ablation study section of our paper. However, it provides valuable insights into the cooperation abilities of agents. The \texttt{Puzzles} maps were specifically designed with narrow corridors, cul-de-sacs, and lacunas. The examples of such scenarios are depicted in Figure~\ref{fig:puzzles}. These challenging instances require agents to cooperate strategically, such as one agent entering a corridor and using a lacuna to let another agent pass.

The results are presented in Table~\ref{tab:results-puzzles}. During testing, there were 2 to 4 agents on each map. Among the learnable approaches, the MAPF-GPT family performed the best, with performance depending on the number of model parameters. The 85M model performs the best for both metrics (success rate and SoC). Interestingly, its success rate is very close to that of LaCAM on some instances. However, LaCAM significantly outperforms all other approaches in terms of SoC. DCC and SCRIMP show much lower results on both metrics. Even for the simplest tasks with two agents, the SoC of DCC and SCRIMP is 2.5 times higher than that of MAPF-GPT-85M and almost 3 times higher than that of LaCAM.

\begin{table}[ht!]
    \centering
    \begin{tabular}{
        l
        S[table-format=1.2(2), separate-uncertainty=true] 
        S[table-format=2.2(4), separate-uncertainty=true]
    }
    \toprule
    \textbf{Algorithm} & \textbf{Success Rate} $\uparrow$ & \textbf{SoC} $\downarrow$ \\
    \midrule
    \multicolumn{3}{c}{Number of agents: 2} \\
    \midrule
    \midrule
    MAPF-GPT-85M    & 1.00 \pm 0.00 & 11.37 \pm 1.27 \\
    MAPF-GPT-6M     & 1.00 \pm 0.00 & 12.69 \pm 2.28 \\
    MAPF-GPT-2M     & 0.99 \pm 0.02 & 16.73 \pm 5.42 \\
    DCC             & 0.91 \pm 0.04 & 31.45 \pm 9.43 \\
    SCRIMP          & 0.95 \pm 0.04 & 28.07 \pm 9.08 \\
    \midrule
    LaCAM           & 1.00 \pm 0.00 & 10.43 \pm 1.07 \\
    \midrule
    \multicolumn{3}{c}{Number of agents: 3} \\
    \midrule
    \midrule
    MAPF-GPT-85M    & 0.99 \pm 0.01 & 27.92 \pm 6.59 \\
    MAPF-GPT-6M     & 0.99 \pm 0.02 & 36.23 \pm 9.73 \\
    MAPF-GPT-2M     & 0.96 \pm 0.03 & 51.07 \pm 14.00 \\
    DCC             & 0.76 \pm 0.07 & 79.92 \pm 14.35 \\
    SCRIMP          & 0.84 \pm 0.06 & 73.97 \pm 18.81 \\
    \midrule
    LaCAM           & 1.00 \pm 0.00 & 19.01 \pm 1.92   \\
    \midrule
    \multicolumn{3}{c}{Number of agents: 4} \\
    \midrule
    \midrule
    MAPF-GPT-85M    & 0.99 \pm 0.02 & 66.30 \pm 12.52 \\
    MAPF-GPT-6M     & 0.91 \pm 0.04 & 101.35 \pm 21.01 \\
    MAPF-GPT-2M     & 0.89 \pm 0.05 & 122.19 \pm 24.64 \\
    DCC             & 0.54 \pm 0.08 & 169.11 \pm 22.41 \\
    SCRIMP          & 0.76 \pm 0.06 & 131.54 \pm 23.82 \\
    \midrule
    LaCAM           & 1.00 \pm 0.00 & 33.09 \pm 3.33 \\
    \bottomrule
    \end{tabular}
    \caption{Comparison of algorithms on \texttt{Puzzles} set, with varying number of agents. $\pm$ shows confidence intervals $95\%$.}
    \label{tab:results-puzzles}
\end{table}

\end{document}